\documentclass[12pt,english,floatfix,superscriptaddress,aps,prd,preprint,showkeys,nofootinbib]{revtex4}
\usepackage{amsmath}
\usepackage{amssymb}
\usepackage{amsbsy}
\usepackage{amsfonts}
\usepackage{amsopn}
\usepackage{amstext}
\usepackage{graphicx}
\usepackage[english]{babel}
\usepackage{color}
\usepackage{slashed}
\usepackage{esint}
\usepackage[dvips]{epsfig}
\usepackage[dvips]{graphicx}
\usepackage{float}
\usepackage{units}
\usepackage{textcomp}
\usepackage{wasysym}
\usepackage{hyperref}
\usepackage{slashed}

\begin{document}

\title{Global monopole in a Ricci-coupled Kalb-Ramond bumblebee gravity}

\author{Fernando M. Belchior}
\email{belchior@fisica.ufc.br}
\affiliation{Departamento de F\'isica, Universidade Federal do Cear\'a,\\ Campus do Pici, 60455-760, Fortaleza, Cear\'a, Brazil.}

\author{Roberto V. Maluf}
\email{r.v.maluf@fisica.ufc.br}
\affiliation{Departamento de F\'isica, Universidade Federal do Cear\'a,\\ Campus do Pici, 60455-760, Fortaleza, Cear\'a, Brazil.}

\author{Albert Yu. Petrov}
\email{petrov@fisica.ufpb.br}
\affiliation{Departamento de F\'isica, Universidade Federal da Para\'iba,
 58051-970, João Pessoa, Para\'iba, Brazil}

\author{Paulo J. Porf\'irio}
\email{pporfirio@fisica.ufpb.br}
\affiliation{Departamento de F\'isica, Universidade Federal da Para\'iba,
 58051-970, João Pessoa, Para\'iba, Brazil}

\begin{abstract}

In this paper, we investigate black hole solutions in Einstein-Kalb-Ramond (EKR) bumblebee gravity sourced by a global monopole characterized by the charge $\eta$. This modified theory of gravity possesses the notable feature of incorporating local Lorentz symmetry breaking (LSB) via a spontaneous symmetry-breaking mechanism.  We solve the field equations for a static and spherically symmetric metric with the Kalb-Ramond (KR) field fixed at its VEV, thereby obtaining new black hole solutions. These solutions simultaneously exhibit the LSB effects, codified by the $\gamma$ parameter, and the global monopole effects, codified by the charge $\eta$. Next, we study the impact of the global monopole and LSB corrections on two classical tests, namely, the advance of Mercury's perihelion and the light deflection. Furthermore, we compute the Hawking temperature, black hole shadows, and greybody factors. Ultimately, we investigate the processes of scattering and absorption of a massless scalar field by employing the partial wave method.

\end{abstract}

\maketitle

\section{Introduction}
The first known black hole (BH) solution of general relativity (GR), proposed by Karl Schwarzschild in 1916 \cite{Schwarzschild}, represents a massive object whose gravitational field is so strong that even light is not able to escape from it. Such solutions are characterized by the presence of singularities, that is, infinite values of geometric invariants. There is a belief that these singularities arise only due to the lack of a quantum description of gravity, see e.g. \cite{Hawking:1973uf}. 
In addition to the Schwarzschild BH, which represents a vacuum solution of the Einstein equations, some other interesting BH solutions were found, such as, for instance, the Reissner–Nordstr\"{o}m metric, corresponding to a charged static black hole \cite{Reissner,Nordstrom}, and Kerr solution corresponding to a rotating BH \cite{Kerr}. Further, In the late 1960s, James Bardeen proposed a (3+1) regular black hole solution that does not exhibit divergences in geometrical invariants \cite{Bardeen}. 

Among other space-times, one can highlight the geometry of the space-time generated by a global monopole (GM) \cite{BARRIOLA,Rhie:1990kc,Dadhich:1997mh}, being described by a field theory with a triplet of interacting scalar fields. The GM solution in GR was obtained for the first time by Barriola and Vilenkin \cite{BARRIOLA}, regarding the region outside the GM core and showing that this geometry possesses a deficit solid angle. In that paper, they also calculated the impact of the GM spacetime on the propagation of null geodesics, showing that light suffers an angular deflection even when one neglects the mass term of the GM. The spacetime generated by a global monopole has been investigated in modified gravity models, such as $f(R)$ gravity in the metric formalism \cite{Man:2013sf,Li:2021ypw}, $f(R)$ gravity in the Palatini formalism \cite{Nascimento:2018sir}, Eddington-inspired Born-Infeld gravity \cite{Nascimento:2019qor,Soares:2023err} and bumblebee gravity \cite{Gullu:2020qzu}.

The study of absorption and scattering of particles due to black holes has been a key issue  in recent studies \cite{Anacleto:2020zhp, Anacleto:2019tdj,Pitelli:2017bgx, Anacleto:2017kmg, Anacleto:2022shk}. Such a study allows us to understand fundamental problems, such as formation of a black hole, as well as its stability and gravitational waves emission. An important result obtained in these papers is that the differential scattering cross section for small angles is given by $d\sigma/d\Omega=16G^2M^2/\theta^4$ at the long wavelength limit, i.e, for $GM\omega<<1$.

The Lorentz symmetry is regarded as a cornerstone of modern physics, playing a pivotal role in GR and the Standard Model of elementary particles.  Nevertheless, there have been many research works addressing the limits of validity of Lorentz symmetry and the possibility of Lorentz symmetry breaking (LSB). This topic gained more prominence after the seminal paper \cite{Kostelecky:1988zi}. In this paper, the authors sought small deviations from Lorentz symmetry in the context of string theory.

The LSB has been implemented in the field theory context through Standard Model Extension (SME), that is, an effective field theory that can describe the standard model coupled to GR incorporating all possible coefficients for violation of Lorentz and/or CPT symmetries \cite{Colladay:1996iz,Colladay:1998fq,Carroll:1989vb}. A significant feature of SME is that the gauge symmetry holds intact. A simple Lorentz violating (LV) model can be constructed with inclusion of a self-interacting vector field called the bumblebee field. Originally, the bumblebee theory has been introduced, both for flat and curved space-times, in \cite{KosGra}. In practical terms, one can write the Lagrangian density for the bumblebee model with a Maxwell-like kinetic term and a potential as written below \cite{KosGra}:
\begin{align}
\mathcal{L}=-\frac{1}{4}\mathcal{B}_{\mu\nu}\mathcal{B}^{\mu\nu}-V(B_\mu B^\mu - b^2), 
\end{align}
where $B_\mu$ is the bumblebee field, $\mathcal{B}_{\mu\nu}=\partial_\mu B_\nu-\partial_\nu B_\mu$ is its field strength and $V(B_\mu B^\mu - b^2)$ is the potential that drives to the spontaneous LSB. Particular forms for that potential have been proposed \cite{Altschul:2005mu,Bluhm:2004ep,Bluhm:2007bd}. Some interesting results for the bumblebee model, especially within studying its perturbative aspects, have been obtained in a flat spacetime, see e. g. \cite{Maluf:2015hda,Belchior:2023cbl}.  At the same time, the most attractive applications of the bumblebee model can be performed in a curved spacetime, where this theory represents itself as the most natural manner to implement the LSB. In recent years, a considerable number of problems of physical interest have been explored within this model, where one can highlight black holes \cite{Maluf:2020kgf,Casana:2017jkc,Guo:2023nkd,Li:2020dln,Ding:2019mal}, wormholes \cite{Ovgun:2018xys,Oliveira:2018oha}, cosmology \cite{Maluf:2021lwh}, matter scattering \cite{Maluf:2014dpa, Maluf:2013nva}, metric-affine formalism \cite{Delhom:2022xfo,Delhom:2019wcm,Delhom:2020gfv,Filho:2022yrk,AraujoFilho:2024ykw} and radiative corrections \cite{Maluf:2015hda,Belchior:2023cbl,Delhom:2020gfv}. On the other hand, one can naturally extend the bumblebee theory to include an antisymmetric tensor field known as the Kalb-Ramond field \cite{Altschul:2009ae,Maluf:2018jwc,Assuncao:2019azw,Aashish:2019ykb,Maluf:2021ywn,Maluf:2021eyu,Lessa:2019bgi,Yang:2023wtu,Fathi:2025byw,Baruah:2025ifh,Liu:2024oas,Liu:2024lve,Filho:2023ycx,AraujoFilho:2024rcr,AraujoFilho:2024ctw,AraujoFilho:2025hkm,Junior:2024vdk,Duan:2023gng}.

In this work, we are interested in obtaining exact black hole solutions with the global monopole in LV Kalb-Ramond gravity both in the absence and presence of cosmological constant. Once we have the solutions in hands, one can study the effects of such geometries by analyzing the geodesics of test particles. Specially, we analyze time-like and null geodesics, corresponding, respectively, to massive particles and photons. Moreover, we employ the partial wave approach to calculate the scattering and absorption cross section of massless scalar particles within obtained geometries.

We outline this present paper as follows. In Section (\ref{s2}), we present a quick review of Einstein-Kalb-Ramond gravity theory, where we derive the equations of motion. Section (\ref{s3}) is devoted to obtaining the black hole solution with and without the cosmological constant. Subsequently, we carry out a study of geodesics in Section (\ref{s4}). The differential scattering and absorption cross-section is analyzed by applying the partial wave method at low- and high-frequency limits in  the Section (\ref{s5}). Finally, our conclusions are summarized in Section (\ref{s6}).


\section{Kalb-Ramond gravity with LSB}\label{s2}

Modified gravity models with LV background fields have become a fruitful line of research. This section discusses the main aspects of a modified gravity theory in which the LV Kalb-Ramond (KR) field couples to the Ricci tensor. This model can be seen as a natural extension of the Einstein–bumblebee gravity originally formulated in \cite{KosGra}. As a starting point, one assumes the gravitational action \cite{Lessa:2019bgi, Yang:2023wtu}
\begin{align}\label{actionKR}
    S=\int d^4x\sqrt{-g}\bigg[\frac{1}{2\kappa}\bigg(R-2\Lambda+\varepsilon\, B^{\mu\lambda}B^\nu\, _\lambda R_{\mu\nu}\bigg)-\frac{1}{12}H_{\lambda\mu\nu}H^{\lambda\mu\nu}-V(B_{\mu\nu}B^{\mu\nu}\pm b^2)+\mathcal{L}_m\bigg],
\end{align}
where $\kappa=8\pi G_N$, being $G_N$ the Newtonian gravitational constant, $\Lambda$ is the cosmological constant and $\varepsilon$ represents the coupling constant between the Ricci tensor and the KR field $B_{\mu\nu}$. The mass dimension of this coupling constant in natural units is $[\varepsilon]=M^{-2}$. In addition, the field strength of $B_{\mu\nu}$ is defined as $H_{\lambda\mu\nu}=\partial_\lambda B_{\mu\nu}+\partial_\mu B_{\nu\lambda}+\partial_\nu B_{\lambda\mu}$. One points out that, as must be for a bumblebee model, the action lacks the gauge symmetry due to nonminimal coupling and the potential $V(B_{\mu\nu}B^{\mu\nu}\pm b^2)$, which is a smooth one that triggers the spontaneous LSB leading to non-vanishing VEV for KR field, i.e., $\langle B_{\mu\nu}\rangle=b_{\mu\nu}$. This potential also induces a non-vanishing VEV associated to dual tensor field $\widetilde{B}_{\mu\nu}$, so that we can write
\begin{align}
 \langle  \widetilde{B}_{\mu\nu} \rangle=\frac{1}{2}\langle \epsilon_{\mu\nu\alpha\beta} B^{\alpha\beta}\rangle=\frac{1}{2}\epsilon_{\mu\nu\alpha\beta} b^{\alpha\beta},
\end{align}
where $\epsilon_{\mu\nu\alpha\beta}$ stands for totally antisymmetric Levi-Civita tensor. We can vary the action (\ref{actionKR}) with respect to metric in order to obtain the gravitational equation
\begin{align}\label{Einstein}
    R_{\mu\nu}+\Lambda g_{\mu\nu}-\frac{1}{2}g_{\mu\nu}R=\kappa(T_{\mu\nu}^m+T_{\mu\nu}^B)+T_{\mu\nu}^{\varepsilon},
\end{align}
where $T_{\mu\nu}^m$ is the energy-momentum tensor associated to the usual matter
\begin{align}
  T_{\mu\nu}^m=-\frac{2}{\sqrt{-g}}\frac{\delta (\sqrt{-g}\mathcal{L}_m)}{\delta g^{\mu\nu}}.  
\end{align}
At the same time, the energy-momentum tensor concerning the pure bumblebee sector is given by
\begin{align}
 T_{\mu\nu}^B=\frac{1}{2}H_{\alpha\beta\mu} H_\nu^{\alpha\beta} -\frac{1}{12}g_{\mu\nu}H_{\lambda\alpha\beta} H^{\lambda\alpha\beta}-g_{\mu\nu}V+4B_{\alpha\mu}B^\alpha _\nu V^{\prime},
\end{align}
while the non-minimal coupling contributes with the energy-momentum tensor 
\begin{align}
 T_{\mu\nu}^{\varepsilon}=\frac{\varepsilon}{\kappa}\bigg(\frac{1}{2}g_{\mu\nu}B^{\alpha\lambda}B^\beta_\lambda R_{\alpha\beta}-B_{\alpha\mu}B_{\beta\nu}R^{\alpha\beta}-B_{\alpha\beta}B_\mu^\alpha R_\nu^\beta-B_{\alpha\beta}B_\nu^\alpha R_\mu^\beta\nonumber\\+\frac{1}{2}\nabla^\alpha\nabla_\mu B_\nu^\beta R_{\alpha}+\frac{1}{2}\nabla^\alpha\nabla_\nu B_\mu^\beta R_{\alpha\beta}-\frac{1}{2}\square B^\alpha_\mu B_{\alpha\nu}-\frac{1}{2}g_{\mu\nu}\nabla_\alpha\nabla_\beta B^{\alpha\lambda}B^\beta_\lambda\bigg).   
\end{align}

On the other hand, if we vary the action with respect to field $B_{\mu\nu}$, we arrive at the following equation of motion:
\begin{align}
    \nabla_\lambda H^{\lambda\mu\nu}=4V^{\prime} B^{\mu\nu}+\frac{\varepsilon}{\kappa} (B^{\mu\lambda} R^\nu\ _\lambda-B^{\nu\lambda} R^\mu\ _\lambda).
\end{align}
The prime above is for the derivative with respect to the argument of the potential. We would like to note that we are not considering any coupling between the matter fields and the background KR field. At the same time, other geometrical couplings could be considered, such as $\varepsilon B^{\mu\nu}B_{\mu\nu}R$ and $\varepsilon  B^{\mu\nu}B^{\alpha\beta}R_{\mu\nu\alpha\beta}$. In this work, we are only interested in the coupling between the KR field and the Ricci tensor. As we will see in the next section, it is convenient to rewrite the Eq. (\ref{Einstein}) as follows
\begin{align}\label{ER}
R_{\mu\nu}-\Lambda g_{\mu\nu}=T_{\mu\nu}-\frac{1}{2}g_{\mu\nu}T,    
\end{align}
where $T_{\mu\nu}=\kappa(T_{\mu\nu}^m+T_{\mu\nu}^B)+T_{\mu\nu}^{\varepsilon}$ is the total energy-momentum tensor and $T$ stands for the trace of tensor $T_{\mu\nu}$. It is noteworthy that this total energy-momentum tensor is conserved due to the Bianchi identities.

\section{Black hole solution with global monopole}\label{s3}
In this section, we will obtain a black hole solution with a global monopole. A basic model to describe a global monopole involves a triplet of scalar fields with a global $O(3)$ symmetry that is spontaneously broken into $U(1)$ symmetry. It is worth mentioning that the gravitational field of a global monopole defines a spacetime with a solid angle deficit. Thereby, the Lagrangian density that describes the global monopole reads \cite{Rhie:1990kc,Dadhich:1997mh}
\begin{align}
\mathcal{L}_m=-\frac{1}{2}\partial_\mu\psi^a\partial^\mu\psi^a - \frac{\chi}{4}(\psi^a\psi^a-\eta^2)^2. 
\end{align}
Above, the index $a$ indicates the labels of the scalar fields $\psi^a$
and runs from $1$ to $3$. Moreover, the parameters $\chi$ and $\eta$ represent the coupling constant and the energy scale in which the symmetry is broken. To proceed, we introduce the following ansatz for $\psi^a$ describing the monopole: $\psi^a=\eta\frac{x^a}{r}$, where $x^ax^a=r^2$. It is worth mentioning that such an ansatz is approximated to outside the GM core. Since our focus is to investigate a static and spherically symmetric spacetime under the nonzero VEV of the KR field, let us assume metric as
\begin{align}\label{metric}
    ds^2=-F(r)dt^2+\frac{dr^2}{F(r)}+r^2(d\theta^2+\sin^2{\theta}d\phi^2).
\end{align}

For the global monopole, the energy–momentum tensor $T^{m}_{\mu\nu}$ is given by \cite{Rhie:1990kc,Dadhich:1997mh}
\begin{align}
T^{m}_{\mu\nu}=\mathrm{diag}\bigg(F(r)\,\frac{\eta^2}{r^2},-\frac{1}{F(r)}\,\frac{\eta^2}{r^2},0,0\bigg).   
\end{align}

Once we have defined the matter sector, we should configure the LV KR field. For this purpose, let us consider a pseudo-electric configuration in which the field $B_{\mu\nu}$ is frozen to its VEV $b_{\mu\nu}$, so that we write its explicit form as \cite{Lessa:2019bgi, Yang:2023wtu}:
\begin{align}\label{KRconfig}
b_{\mu\nu}=b_{01}=-b_{10}=\frac{\vert b \vert}{\sqrt{2}}.    
\end{align}

This configuration leads to the constant norm $b_{\mu\nu}b^{\mu\nu}=-\vert b \vert^2$, and $X= B^{\mu\nu} B_{\mu\nu}+b^2$. It is easy to verify that the field strength $H_{\lambda\mu\nu}$  identically vanishes. Then, the equations of motion for $B_{\mu\nu}$ are satisfied. Initially, we will focus on solution without cosmological constant. In this case, we adopt a quartic potential, i.e., $V(X)=\lambda X^2$, where $\lambda$ is a coupling constant, and $X=B_{\mu\nu}B^{\mu\nu}-b^2$. For this potential, the configuration (\ref{KRconfig}) implies $V=V'=0$.

With the metric ansatz (\ref{metric}) and the aforementioned considerations, we can write the gravitational equations (\ref{ER}) explicitly:
\begin{align}
r F''(r)+2 F'(r)=0,    
\end{align}
and
\begin{align}
 \gamma\, r^2 F''(r)+\left(1+\gamma\right)r F'(r)+ F(r) \left(1-\gamma\right)-(1-\kappa\eta ^2)=0,    
\end{align}
where we have defined the dimensionless LV parameter $\gamma=\varepsilon\,\vert b\vert^2 /2$.  Some manipulations lead to the equation
\begin{align}
 r \left(1-\gamma\right) F'(r)+ F(r) \left(1-\gamma\right)+\kappa\eta ^2-1=0,
\end{align}
whose solution is
\begin{align}\label{sol1}
    F(r)=\frac{1-\kappa\eta^2}{1-\gamma}-\frac{2M}{r}.
\end{align}
This solution differs from that obtained in \cite{Fathi:2025byw, Baruah:2025ifh} since it was considered a different Lagrangian.

It should be noted that the parameter $\gamma$ represents the magnitude of Lorentz symmetry violation, induced by the non–zero VEV of the KR field. When $\gamma=\eta=0$, one recovers the Schwarzschild solution. Moreover, the Kretschmann scalar for this geometry is given by
\begin{align}
 R_{\mu\nu\alpha\beta} R^{\mu\nu\alpha\beta}=\frac{48 M^2}{r^6}+\frac{16M(\kappa\eta^2-\gamma)}{r^5(1-\gamma)}-\frac{4(\kappa^2\eta^4+2\gamma\kappa\eta^2-\gamma^2)}{r^4(1-\gamma)^2}. 
\end{align}

As we can see, in this case, LV effects cannot be eliminated by a simple coordinate transformation. One notes that there is unique Schwarzschild-like singularity localized at $r=0$. Besides, the solution (\ref{sol1}) possesses a horizon with the radius
\begin{align}
r_h=\frac{2M(1-\gamma)}{1-\kappa\eta^2}.    
\end{align}
This horizon is shifted by the LV parameter. Additionally, the Hawking temperature reads
\begin{align}
 T_H=\frac{1}{8\pi} \frac{(1-\kappa\eta^2)^2}{(1-\gamma^2)^2}. 
\end{align}

On the other hand, when the cosmological constant is assumed to be non-zero, we realize
that the assumption in which the KR field is frozen in its VEV, i.e., $V=0$, does not support a suitable solution that satisfies the equations of motion. Therefore, we need to follow an alternative approach that can be accomplished by using a linear potential, i.e., $V(X)=\tilde{\lambda} X$, with $\tilde{\lambda}$ being now a Lagrange multiplier field (see f.e. \cite{Maluf:2020kgf,Yang:2023wtu}). It is easy to see that the derivative of this potential with respect to X is $V_X(X)=\tilde{\lambda}$. Thus, when $\Lambda\neq 0$, we write the gravitational equations (\ref{ER}) explicitly below 
\begin{align}\label{ee1}
  (\gamma -1) r F''+2 (\gamma -1) F'-2 \Lambda  r=0,   
\end{align}
and
\begin{align}\label{ee2}
\gamma  r^2 F''+(\gamma +1) r F'-(\gamma -1) F+r^2 \left(b^2 \lambda +\Lambda \right)+\kappa\eta ^2 -1=0.    
\end{align}
After some manipulations with Eqs. (\ref{ee1},\ref{ee2}), we arrive at
\begin{align}
r(1-\gamma) F'-(\gamma -1) F+r^2 \left(b^2 \lambda +\frac{(1-3 \gamma) \Lambda }{1-\gamma}\right)+\kappa\eta ^2 \kappa -1=0.   
\end{align}

We can solve the above equation analytically. Its solution is given by 
\begin{align}\label{fwc}
F(r)=\frac{1-\kappa\eta^2}{1-\gamma}-\frac{(1-3\gamma)\Lambda+(1-\gamma)\tilde{\lambda} b^2)}{3(1-\gamma)^2}r^2-\frac{2M}{r}.    
\end{align}
By substituting (\ref{fwc}) into (\ref{ee1}), we can show that the on-shell value of $\tilde{\lambda}$ is determined by
\begin{align}
 \lambda=\frac{3\gamma\Lambda}{(1-\gamma)b^2}.   
\end{align}
Thereby, the black hole solution reads
\begin{align}\label{sol2}
F(r)=\frac{1-\kappa\eta^2}{1-\gamma}-\frac{\Lambda}{3(1-\gamma)}r^2-\frac{2M}{r}.
\end{align}
It is straightforward to see that the solution (\ref{sol2}) reduces to (\ref{sol1}) when the cosmological constant vanishes. We can still affirm that this solution is compatible with the equations of motion. At infinity, the solution (\ref{sol2}) tends to the asymptotical value $-\frac{\Lambda}{3(1-\gamma)}r^2$, showing the same asymptotic behavior as the $(A)dS$ metric. In this case, the Kretschmann scalar looks like
\begin{align}
 R_{\mu\nu\alpha\beta} R^{\mu\nu\alpha\beta}=\frac{48 M^2}{r^6}+\frac{16 M \left(\kappa\eta ^2-\gamma\right)}{(1-\gamma) r^5}+\frac{4 \left(\kappa\eta ^2-\gamma\right)^2}{(1-\gamma)^2 r^4}+\frac{8 \Lambda  \left(\kappa\eta ^2-\gamma\right)}{3 (1-\gamma)^2 r^2}+\frac{8 \Lambda ^2}{3 (1-\gamma)^2}.  
\end{align}
The physical horizon for this geometry is given by
\begin{align}
    r_h=-\frac{\left(1-\kappa\eta ^2\right) (1-\gamma)}{\Pi(M,\Lambda,\gamma)^{1/3}}-\frac{\Pi(M,\Lambda,\gamma)^{1/3}}{\Lambda},
\end{align}
where we have defined
\begin{align}\label{lvp}
\Pi(M,\Lambda,\gamma)=\sqrt{\Lambda ^3 (1-\gamma)^3 \left(9 \Lambda  (1-\gamma) M^2-\left(1-\kappa\eta ^2\right)^3\right)}-3 \Lambda ^2 (1-\gamma)^2 M.    
\end{align}

Additionally, one can find that the Hawking temperature is
\begin{align}
    T_H=\frac{\Lambda ^2 M\, \Pi(M,\Lambda,\gamma)^{2/3}}{2 \pi  \left[\Pi(M,\Lambda,\gamma)^{2/3}+\Lambda\left(1-\kappa\eta ^2\right) (1-\gamma)\right]^2}.
\end{align}

Using (\ref{lvp}), one can see that when the cosmological constant is negative ($AdS$ case), there are acceptable black hole solutions for any parameter $\gamma$. On the other hand, when the cosmological constant is positive ($dS$ case), the black hole solutions exist only when the parameters obey the relation $9 \Lambda  (1-\gamma) M^2\geq\left(1-\kappa\eta ^2\right)^3$. When $9 \Lambda  (1-\gamma) M^2=\left(1-\kappa\eta ^2\right)^3$, one faces the situation in which the event horizon and the cosmological horizon are equivalent.


\section{Geodesics}\label{s4}
Having obtained the exact solution for the geometry, the next step is to investigate its properties by analyzing the geodesics associated with this spacetime. The spherical symmetry allows us to restrict the consideration by the radial geodesics. Further, we shall show possible situations in which the spacetime is geodesically complete. We will consider the effects of the modified geometry on the circular orbits of test particles.
\begin{align}
  \mathcal{L}=\frac{1}{2}g_{\alpha\beta}\frac{d x^\alpha}{d\tau}\frac{d x^\beta}{d\tau}. 
\end{align}

Let us consider the equatorial plane $\theta = \pi/2$, so that the metric (\ref{metric}), with $F(r)$ given by (\ref{sol1}) provides, without loss of generality, the following expression for the particle Lagrangian
\begin{align}\label{eqg}
\mathcal{L}=-F(r)\Dot{t}^2+\frac{1}{F(r)}\Dot{r}^2+r^2\Dot{\phi}^2,
\end{align}
where the dot represents derivative with respect to an affine parameter denoted by $\tau$. 
In addition, $\mathcal{L}$ takes the values $-1/2$ (time-like geodesics), $0$ (null geodesics), and $1/2$ (space-like geodesics). It is worth highlighting that the considered metric has two Killing vector fields, namely $\partial/\partial t$ and $\partial/\partial \phi$. The first Killing vector is associated with energy conservation, whereas the second one is associated with the conservation of the angular momentum. The conserved energy $E$ and angular momentum $L$ for a particle along the geodesics read
\begin{align}
\label{Ec} E=-\frac{\partial\mathcal{L}}{\partial\Dot{t}}=F(r)\Dot{t},\\
\label{Lc} L=\frac{\partial\mathcal{L}}{\partial\Dot{\phi}}=r^2\Dot{\phi}.
\end{align}
Then, by putting (\ref{Ec}) and (\ref{Lc}) into (\ref{eqg}), we derive the radial geodesic, namely
\begin{align}\label{me}
\Dot{r}^2=E^2-2F(r)\bigg(\frac{L^2}{2r^2}-\mathcal{L}\bigg),    
\end{align}
which describes the motion of a point particle. Such a equation can be cast in the form
\begin{align}
\frac{1}{2}\left(\frac{d r}{d\tau}\right)^2+\mathcal{U}_{g}=\mathcal{E},  
\end{align}
so that we are able to define the gravitational potential $\mathcal{U}_{g}$ and the energy $\mathcal{E}$ as follows:
\begin{align}
\mathcal{U}_{g}=F(r)\left(\frac{L^2}{2r^2}-\mathcal{L}\right),    
\end{align}
and
\begin{align}
\mathcal{E}=\frac{E^2}{2}.    
\end{align}
With these expressions in hand, it is possible to examine the motion of test particles. In this sense, it is straightforward to observe that the regions where $\mathcal{E}=\mathcal{U}_{g}$ lead to turning points. On the other hand, when $\mathcal{E}-\mathcal{U}_{g}>0$, the radial motion of massive particles takes place. In addition, the circular orbits are possible when the effective potential yields the condition $\frac{d \mathcal{U}_{g}}{dr}=0$ and such orbits are stable when the requirement $\frac{d^2 \mathcal{U}_{g}}{dr^2}>0$ is satisfied. Further, we shall investigate the impact of LSB on geodesics for massive and massless test particles.

\subsection{Time-like particle and the advance of the Mercury perihelion}
First, let us turn our attention to massive particles, whose geodesics are time-like. In this conjecture, we can study the precession of
Mercury perihelion. Then, by taking $\mathcal{L}=-1/2$ in (\ref{me}), which corresponds to massive objects, one obtains
\begin{align}\label{me12}
\Dot{r}^2=E^2-F(r)\bigg(\frac{L^2}{r^2}+1\bigg).    
\end{align}
Now we can write the radial coordinate $r$ in terms of the angle $\phi$. With help of (\ref{Lc}) and using (\ref{sol1}), we rewrite (\ref{me12}) as follows
\begin{align}\label{rf}
\bigg(\frac{dr}{d\phi}\bigg)^2=r^4\bigg(\frac{E^2}{L^2}\bigg)-\frac{r^4}{L^2}\bigg(\frac{1-\kappa\eta^2}{1-\gamma}-\frac{2M}{r}\bigg)\bigg(1+\frac{L^2}{r^2}\bigg).   
\end{align}
At this point, it is convenient to introduce a new coordinate, $u=\frac{L^2}{Mr}$. After this change of coordinates, Eq. \eqref{rf} becomes
\begin{equation}
\bigg(\frac{du}{d\phi}\bigg)^2+\bigg[\frac{L^2}{M^2}\bigg(\frac{1-\kappa\eta^2}{1-\gamma}\bigg)+\bigg(\frac{1-\kappa\eta^2}{1-\gamma}\bigg) u^2-2u-\frac{2M^2}{L^2}u^3
\bigg]=\frac{E^2L^2}{M^2}.
\label{rf2}
\end{equation}
 Now, we should differentiate Eq. (\ref{rf2}) with respect to $\phi$ to obtain a second-order differential equation. Thus, one can write
\begin{align}
\frac{d^2u}{d\phi^2}+\bigg(\frac{1-\kappa\eta^2}{1-\gamma}\bigg)u-1-\frac{3M^2}{L^2}u^2=0.
\label{u}
\end{align}
It should be noted that if we turned off the LV and monopole effects, i.e., by setting $\eta=\gamma=0$, one recovers the standard GR results. A simple way to address this equation is using a perturbative scheme. Namely, the general solution can be expanded as follows:
\begin{align}
u=u_0+u_1+\cdots,
\label{pert0}
\end{align}
where $u_0$ represents the unperturbed solution, while $u_1$ is the first–order perturbed solution. The ellipses represent higher–order corrections, which will be disregarded within this perturbative approach. Applying this methodology, Eq. \eqref{u} can be solved iteratively. Therefore, by straightforwardly substituting Eq.\eqref{pert0} into Eq. \eqref{u}, one finds that the zeroth-order solution must fulfill the following equation,
\begin{equation}
\frac{d^2u_0}{d\phi^2}+\bigg(\frac{1-\kappa\eta^2}{1-\gamma}\bigg)u_0-1=0.
\label{u0}
\end{equation}
 Similarly, the first-order solution must satisfy,
 \begin{equation}
 \frac{d^2u_1}{d\phi^2}+\bigg(\frac{1-\kappa\eta^2}{1-\gamma}\bigg)u_1 -1=\frac{3M^2}{L^2}u^2_0.
 \label{u1}
 \end{equation}
Solving Eq. \eqref{u0}, we get the zeroth-order solution 
\begin{equation}
u_0=\frac{1-\gamma}{1+\kappa\eta^2}\left(1+\epsilon\cos\bigg(\sqrt{\frac{1-\kappa\eta^2}{1-\gamma}}\phi\bigg)\right).
\end{equation}
Now, plugging it into Eq. \eqref{u1}, one finds the first-order solution
\begin{eqnarray}
\nonumber u_{1}&=&\frac{3M^2}{L^2}\bigg[\bigg(\frac{1-\gamma}{1-\kappa\eta^2}\bigg)^{3}\left(1+\frac{1}{2}\epsilon^2\right)-\frac{1}{6}\bigg(\frac{1-\gamma}{1-\kappa\eta^2}\bigg)^{3}\epsilon^2 \cos\bigg(\sqrt{\frac{1-\kappa\eta^2}{1-\gamma}}\phi\bigg)+\\
&+&\bigg(\frac{1-\gamma}{1-\kappa\eta^2}\bigg)^{5/2}\epsilon\phi\sin\bigg(\sqrt{\frac{1-\kappa\eta^2}{1-\gamma}}\phi\bigg)\bigg].
\label{u1s}
\end{eqnarray}
The only relevant term in Eq. \eqref{u1s} is the third term, as the other two describe constant and oscillatory terms around zero, which can therefore be neglected. Taking this into account, the complete solution, up to first-order, is given by 
\begin{align}
u=\frac{1-\gamma}{1+\kappa\eta^2}\bigg[1+\epsilon\cos\bigg(\sqrt{\frac{1-\kappa\eta^2}{1-\gamma}}\phi\bigg)+\frac{3M^2}{L^2}\bigg(\frac{1-\gamma}{1-\kappa\eta^2}\bigg)^{3/2}\epsilon\phi\sin\bigg(\sqrt{\frac{1-\kappa\eta^2}{1-\gamma}}\phi\bigg)\bigg],    
\end{align}
where $\epsilon$ is the eccentricity of the orbit. Experimental data from the Solar System show that $\frac{3M^2}{L^2}<<1$, one can cite Mercury as an example.  In this case, the former equation reduces to an elliptical equation, i.e.,
\begin{align}
u=\frac{1-\gamma}{1-\kappa\eta^2}\bigg\{1+\epsilon\cos\bigg[\bigg(1-\frac{3M^2(1-\gamma)^2}{L^2(1-\kappa\eta^2)^2}\bigg)\sqrt{\frac{1-\kappa\eta^2}{1-\gamma}}\phi\bigg]\bigg\}.   
\end{align}
The next step is to define the period of the non–circular orbits, namely
\begin{align}
T=2\pi \sqrt{\frac{1-\gamma}{1-\kappa\eta^2}}\bigg(1-\frac{3M^2(1-\gamma)^2}{L^2(1-\kappa\eta^2)^2}\bigg)^{-1}\approx 2\pi \sqrt{\frac{1-\gamma}{1+\kappa\eta^2}}\bigg(1+\frac{3M^2(1-\gamma)^2}{L^2(1-\kappa\eta^2)^2}\bigg).   
\end{align}
We write the advance of perihelion defined by $\Delta T=T-2\pi$ for each period of the Mercury as follows
\begin{align}
\Delta T=\frac{6\pi M^2}{L^2(1-\kappa\eta^2)^{5/2}}+2\pi \bigg(\frac{1}{\sqrt{1-\kappa\eta^2}}-1\bigg)-\frac{\pi\gamma}{\sqrt{1-\kappa\eta^2}},   
\label{adv}
\end{align}
where we have considered the lowest order in the expansion of $\gamma$. Two first terms represent the estimation for a GM black hole in general relativity, while the third term represents the contribution arising from LV effects.

\subsubsection{Estimation of the symmetry-breaking energy scale of the global monopole based on the advance of Mercury’s perihelion}

Having obtained the expression for the advance of Mercury's perihelion, we aim to estimate the
symmetry-breaking energy scale of the global monopole, $\eta$, from the available observational data. This estimation involves comparing the GR theoretical prediction of the advance of Mercury's perihelion with the experimental data.  Before proceeding further, we assume that the contributions of the global monopole parameter and the LSB coefficient are small at the Solar System level, then Eq. \eqref{adv} can be broken into three contributions, namely,
\begin{equation}
\Delta T=\delta_{GR}T+\delta_{GM}T+\delta_{LSB}T, 
\end{equation}
where $\delta_{GR}T$ is the standard Schwarzschild contribution from GR,  $\delta_{GM}T$ is the contribution due to the global monopole and $\delta_{LSB}T$ is the LSB contribution. Their explicit forms are given by 
\begin{eqnarray}
\delta_{GR}T&=&\frac{6\pi GM}{c^2 (1-\epsilon^2)a};\\
\delta_{GM}T&=&\frac{8\pi^2 G\eta^2}{c^4};\\
\delta_{LSB}T&=&-\pi \gamma,
\end{eqnarray}
where we restored the speed of light $c$ and gravitational constant $G$ and used the definition $L^2=\frac{GM}{c^2}(1-\epsilon^2)a$, with $a$ being the semi-major axis of the orbital ellipse. The theoretical standard GR contribution is computed using the data table \cite{link, table}, $\delta_{GR}T=42.981^{\prime\prime}/\mbox{century}$. Recent experimental data point to a departure from the GR prediction of $-0.002\pm 0.003^{\prime\prime}/\mbox{century}$. Such a discrepancy could indicate the effects of physics beyond GR.  With this at hand, we can estimate a bound for the global monopole-symmetry-breaking energy scale $\eta$ from observational data of the advance of Mercury's perihelion. Since LSB contribution ($\delta_{LSB}T$) has already been exhaustively estimated from several Solar System experiments \cite{Casana:2017jkc, Filho:2022yrk, AraujoFilho:2024ykw}, any extra contribution beyond GR should be smaller than the upper bound value of the uncertainty, i.e., 
\begin{equation}
\delta_{GM}T=\frac{8\pi^2 G\eta^2}{c^4}<0.003^{\prime\prime}/\mbox{century}=72.3\times 10^{-7\,\prime\prime}/\mbox{orbit}=3.5\times 10^{-11}\mbox{rad}/\mbox{orbit}.
\end{equation}
Now, solving this inequality one finds that the global monopole vacuum energy scale is bounded from above, $\eta<8\times 10^{12}\,\mbox{GeV}$. This result is four orders of magnitude smaller than one found by \cite{BARRIOLA}.

\subsection{Light deflection and gravitational lensing}
The next step is to analyze the null geodesics, which are obtained by taking $\mathcal{L}=0$, as it must be for massless objects (we note that the situation is hence different from Subsection IV.A where one had $\mathcal{L}=-1/2$, which corresponds to the massive objects, in particular, planets, considered there). Thus, from (\ref{eqg}) with $\mathcal{L}=0$, the equation for each photon is given by
\begin{align}\label{me0}
\frac{1}{2}\left(\frac{d r}{d\tau}\right)^2+\mathcal{U}_{g}=\mathcal{E},  
\end{align}
where we have defined the gravitational potential $\mathcal{U}_{g}$ and the energy $\mathcal{E}$ for photon as follows:
\begin{align}
\mathcal{U}_{g}=F(r)\left(\frac{L^2}{2r^2}\right),    
\end{align}
and
\begin{align}
\mathcal{E}=\frac{E^2}{2}.    
\end{align}
The next step is to analyze the unstable orbits that satisfy the conditions $\frac{d \mathcal{U}_{g}}{dr}=0$ and $\frac{d^2 \mathcal{U}_{g}}{dr^2}<0$. Initially, let us analyze the case with the zero cosmological constant, where the gravitational potential is
\begin{align}
\label{phU}
\mathcal{U}_{g}=\left(\frac{1-\kappa\eta^2}{1-\gamma}-\frac{2M}{r}\right)\left(\frac{L^2}{2r^2}\right). 
\end{align}
From this potential, we obtain the critical radius:
\begin{align}\label{cr}
  r_c=\frac{3M(1-\gamma)}{1-\kappa\eta^2}.  
\end{align}
Concurrently, the effective gravitational potential for the case with a non-zero cosmological constant looks like
\begin{align}
\mathcal{U}_{g}=\left(\frac{1-\kappa\eta^2}{1-\gamma}-\frac{\Lambda}{3(1-\gamma)^2}r^2-\frac{2M}{r}\right)\left(\frac{L^2}{2r^2}\right).    
\end{align}
This potential provides the same critical radius as above, (\ref{cr}). Let us proceed with investigating the light deflection in the weak field approach. To achieve this goal, we rewrite (\ref{me0}) as follows
\begin{align}\label{rf0}
\bigg(\frac{dr}{d\phi}\bigg)^2=r^4\bigg(\frac{E^2}{L^2}\bigg)-r^2\bigg(\frac{1-\kappa\eta^2}{1-\gamma}-\frac{2M}{r}\bigg),
\end{align}
leading to
\begin{align}\label{ad}
    \frac{d\phi}{dr}=\bigg[\frac{r^4}{\beta^2}-r^2\bigg(\frac{1-\kappa\eta^2}{1-\gamma}-\frac{2M}{r}\bigg)\bigg]^{-1/2},
\end{align}
where we have defined the impact parameter as being $\beta=\frac{L}{E}$. Hence, one denotes the turning point of a particular orbit by $r_0$. Besides, it is admitted that the photon set off from an asymptotically flat region, approaching the black hole and then deviating at the closest distance $r_0$, being $r_0>r_c$, before traveling to another asymptotically flat region of spacetime. Thereby, from (\ref{ad}), we obtain
\begin{align}
\Delta\phi=2\int_{r_0}^{\infty}\bigg[\frac{r^4}{\beta^2}-r^2\bigg(\frac{1-\kappa\eta^2}{1-\gamma}-\frac{2M}{r}\bigg)\bigg]^{-1/2} dr.   
\end{align}
To evaluate the above integral, it is convenient to introduce the new variable $u=\frac{1}{r}$. Thus, the impact parameter becomes
\begin{align}
\beta=\frac{1}{u_0}\bigg(\frac{1-\kappa\eta^2}{1-\gamma}-2Mu_0\bigg)^{-1/2}.    
\end{align}
After some manipulations, we arrive at
\begin{align}\label{lb}
\Delta\phi=2\int_{0}^{u_0}\bigg[u_0^2\bigg(\frac{1-\kappa\eta^2}{1-\gamma}-2Mu_0\bigg)-u^2\bigg(\frac{1-\kappa\eta^2}{1-\gamma}-2Mu\bigg)\bigg]^{-1/2} du.    
\end{align}
We can evaluate (\ref{lb}), taking into account the terms up to the first order in $M$ (weak field limit). Thus, by performing the integration over $u$, we get the deflection of light given by
\begin{align}\label{Df}
\Delta\phi=\pi\sqrt{\frac{1-\gamma}{1-\kappa\eta^2}}+\frac{4M}{\beta}\bigg(\frac{1-\gamma}{1-\kappa\eta^2}\bigg)^2 + \mathcal{O}(M^2).   
\end{align}
To proceed further, one assumes that the LV effects are small, so that we can expand (\ref{Df}) up to the first order in the LV parameter $\gamma$.
Finally, the angular deflection up to the first order can be written up to the first order in $\gamma$ and $M$ as follows
\begin{align}\label{ang}
\delta\phi=\Delta\phi-\pi\approx \pi\bigg[\frac{1-\frac{\gamma}{2}}{\sqrt{1-\kappa\eta^2}}-1\bigg]+\frac{4M}{\beta} \frac{1-2\gamma}{(1-\kappa\eta^2)^2}.  
\end{align}

As one can observe, the first term provides the angular deflection when $M=0$, leading to a non-zero light deflection. On the other hand, the second term represents the angular deflection for the black hole with mass $M$. One still notes that both terms bring contributions of Lorentz symmetry violation. Having the relationship of the deflection angle and the impact parameter, we are able to find the classical differential scattering section at the limit of small $\delta\phi$ through the following equation \cite{Anacleto:2022shk}
\begin{align}
 \frac{d\sigma}{d\Omega}= \frac{\beta}{\delta\phi} \frac{d\beta}{d(\delta\phi)} 
\end{align}
Thus, we obtain explicitly,
\begin{align}
    \frac{d\sigma}{d\theta}=\frac{16 M^2 (1-\gamma)^4}{(1-\kappa\eta^2)^4\theta^4}\approx\frac{16 M^2 (1-4\gamma)}{(1-\kappa\eta^2)^4\theta^4}
\end{align}
Another important result is the absorption, which is determined by
\begin{align}
   \sigma_{abs}=\pi \beta^2=\frac{4\pi\delta_0^2}{\omega^2}=\frac{16\pi M^2 (1-\gamma)^4}{(1-\kappa\eta^2)^4} 
\end{align}
In the next section, we will study the scattering and absorption of a massless scalar field due to the LV black hole solution we have previously analyzed.

\subsubsection{Estimation of the symmetry-breaking energy scale of the global monopole based on the light deflection}

Given the equation for the angular deflection \eqref{ang}, one can use the well-known parameterized post-Newtonian (PPN) approach \cite{Misner, Will} to estimate the symmetry-breaking energy scale from observational data. It is noteworthy that, for our purposes, $\gamma^{\prime}$ is the only relevant PPN parameter\footnote{This parameter is traditionally denoted by $\gamma$; however, since $\gamma$ is already used for the LSB parameter, we denote the PPN parameter as $\gamma^{\prime}$.}. This parameter is responsible for PPN contributions of space–curvature to the deflection of light. As expected, the standard GR results are recovered when $\gamma^{\prime}=1$. Using this approach, it was shown in \cite{Will2, Klioner} that the PPN contribution to the angular deflection of a light ray passing by a massive body at a distance $\beta$ looks like
\begin{align}
    \delta_{_{PN}}\phi=\frac{1}{2}(1+\gamma^{\prime})\frac{4GM}{c^2 \beta}\frac{1+\cos\chi}{2},
    \label{ppn}
\end{align}
where $M$ is the mass of the massive body and $\chi$ is the angle between the massive body and the source. Let us suppose a grazing light ray, corresponding to taking $\beta\approx R_\odot$ and $\chi\approx 0$. In addition, we shall consider the massive body to be the Sun, i.e., $M=M_{\odot}$. Putting all this information together, Eq. \eqref{ppn} reduces to
\begin{align}
    \delta_{_{PN}}\phi= \frac{1}{2}(1+\gamma^{\prime})\,1.75^{\prime\prime}, 
    \label{PN2}
\end{align}
where, for $\gamma=1$, one recovers the GR prediction, $\delta_{_{GR}}\phi=1.75^{\prime\prime}$.
 
Based on this discussion, one can estimate the contributions beyond GR -- which stems from the LSB parameter ($\gamma$) and the symmetry-breaking energy scale ($\eta$) -- from observational data. Since the additional contribution to GR is codified by the PPN parameter $\gamma^{\prime}$, we might use the experimental data on $\gamma^{\prime}$ to estimate $\eta$ (we focus on the calculation of $\eta$ because the LSB parameter has already been estimated in \cite{Filho:2022yrk}). In this regard, it pays to note that several experimental tests constrain the value of $\gamma^{\prime}$ parameter at the Solar System level (see, for example, \cite{Bertotti, Robertson,Lebach,Lambert,Shapiro, Fomalont, Titov}. More recently, the formal error in $\gamma^{\prime}$ was estimated to be $(\gamma^{\prime}-1)=\pm 0.9\times 10^{-4}$ using Very-Long-Baseline Interferometry (VLBI) technology \cite{Titov}. Similarly to the previous subsection,  let us assume that the contributions stemming from the global monopole parameter and the LSB coefficient are small at the Solar System level. In this case, Eq. \eqref{ang} can be split into three pieces,
\begin{align}
\delta\phi=\delta_{_{GR}}\phi+\delta_{_{GM}}\phi+\delta_{_{LSB}}\phi,
 \end{align}
where
\begin{align}
    \delta_{_{GR}}\phi=\frac{4GM_{\odot}}{c^2 R_\odot}=1.75^{\prime\prime}
\end{align}
is the standard GR contribution for the angular deflection,
\begin{align}
    \delta_{_{GM}}\phi=\frac{4\pi^2 G \eta^2}{c^4}+\frac{64G^2 M^2_{\odot}\eta^2}{c^6 R_{\odot}}
\end{align}
is the global monopole contribution for the angular deflection and
\begin{align}
    \delta_{_{LSB}}\phi=-\frac{\pi\gamma}{2}-\frac{8G M_{\odot}\gamma}{c^2 R_{\odot}}
\end{align}
is the LSB contribution for the angular deflection. As said before, we shall focus on estimating the symmetry-breaking energy scale of the global monopole. To do that, we assume that (since global monopoles have not yet been observed in nature) $\delta_{_{GM}}\phi$ is smaller than the uncertainty in the measurement of the PPN parameter, $\gamma^{\prime}$; that is, 
 \begin{align}
 \delta_{_{GM}}\phi< \frac{(\gamma^{\prime}-1)}{2}1.75^{\prime\prime}.
 \end{align}
 Using the data of \cite{Titov}, which estimates $|\gamma^{\prime}-1|= 0.9\times 10^{-4}$, we find the following upper bound for the symmetry-breaking energy scale of the global monopole, $\eta<3.8\times 10^{13}$ GeV. Such a result is one order of magnitude larger than that obtained for the advance of Mercury's perihelion.

\subsubsection{Black hole shadow}

An important point to discuss here is the formation of a black hole shadow. One can comprehend such a phenomenon as the result of the interaction between the strong gravitational field caused by the black hole and the surrounding light ray. Since the black hole deforms space-time, a passing photon is deflected. There are two possibilities for photons: if they have small orbital angular momentum, they are trapped by the black hole. On the other hand, if the photons have large orbital angular momentum, they can escape from it. As a consequence, a distant observer perceives a dark zone in the sky, which is known as the black hole shadow. The radius of the circular photon orbit is known critical radius, given by (\ref{cr}) for the potential (\ref{phU}). This radius defines the photon sphere. 

\begin{figure}[ht!]
\begin{center}
\hspace*{-3mm}
\begin{tabular}{ccc}
\includegraphics[scale=0.242]{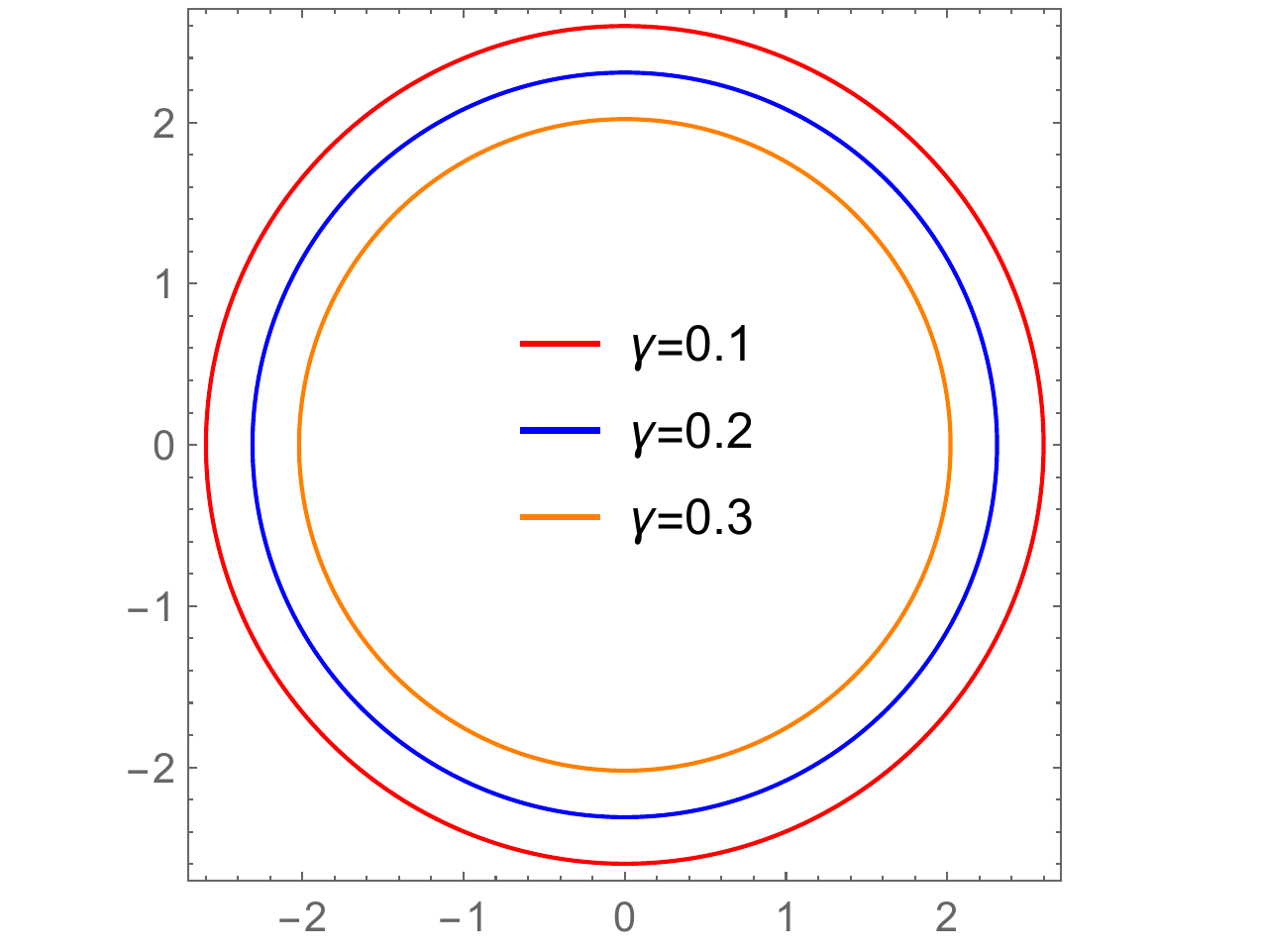} 
\includegraphics[scale=0.242]{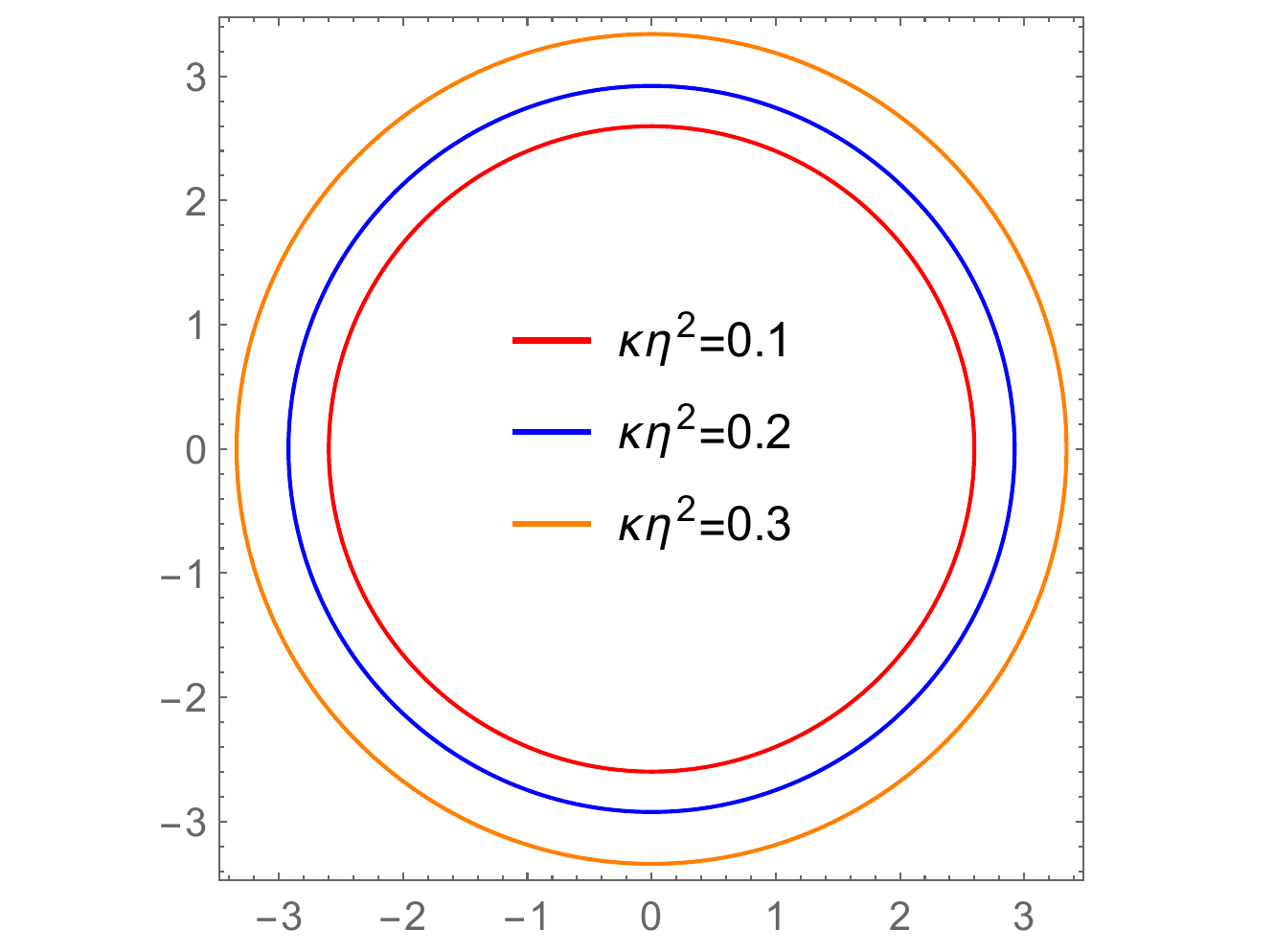}\\
(a)\hspace{8.5cm}(b)\\
\end{tabular}
\end{center}
\vspace{-0.5cm}
\caption{ The circles in the figure illustrate shadows formed by varying the parameters $\gamma$ for $\kappa\eta^2=0.1$ (a) and $\eta$ for $\gamma=0.1$ (b). We have set $M=0.5$.
\label{fig1}}
\end{figure}

\begin{figure}[ht!]
\begin{center}
\hspace*{-3mm}
\begin{tabular}{ccc}
\includegraphics[scale=0.23]{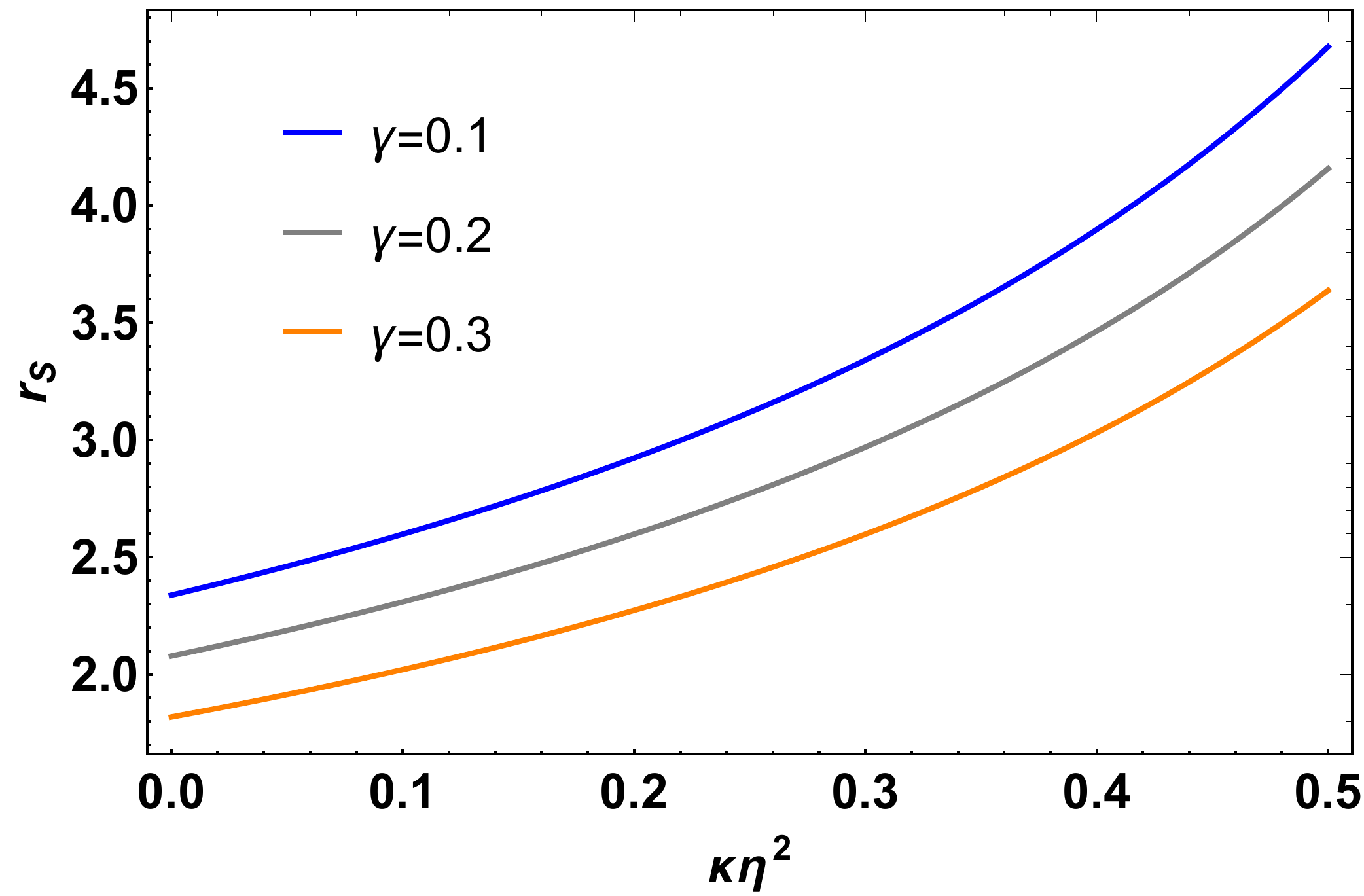} 
\includegraphics[scale=0.23]{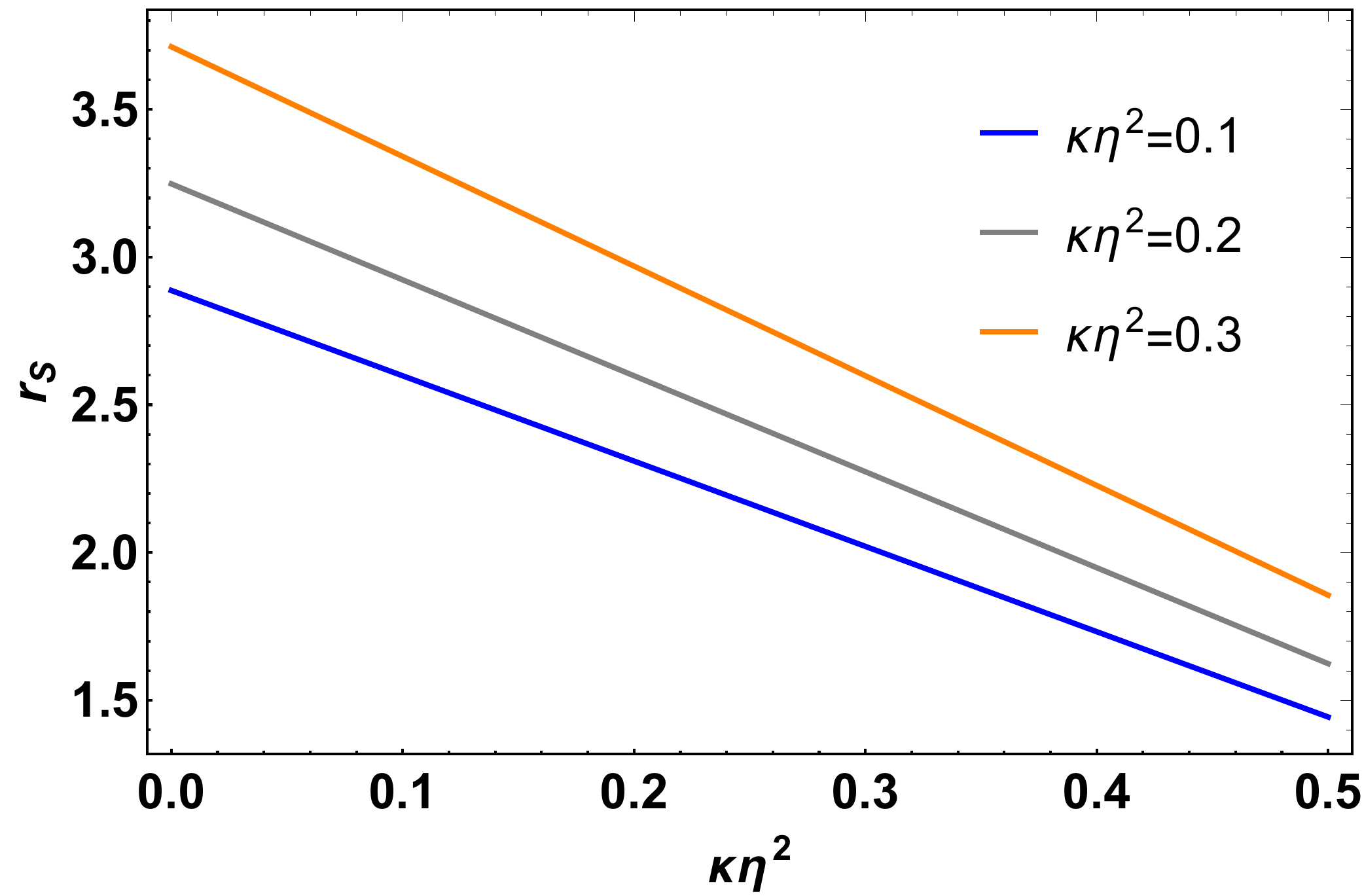}\\
(a)\hspace{8.5cm}(b)\\
\end{tabular}
\end{center}
\vspace{-0.5cm}
\caption{Shadow radius versus parameter the parameter $\gamma$ (a) and versus the parameter $\eta$ (b). We have set $M=0.5$.
\label{fig2}}
\end{figure}

With the critical radius, we can calculate the size of the black hole shadow. Thus, a static observer at the position $r_0$ sees a shadow, whose radius is given by
\begin{align}\label{sr}
  r_s=\sqrt{\frac{F(r_0)}{F(r_c)}}r_c=\frac{3M(1-\gamma)}{1-\kappa\eta^2}\sqrt{3-\frac{6M(1-\gamma)}{r_0(1-\kappa\eta^2)}}. \end{align}
For the observer located at infinity, we get an explicit formula for the shadow radius, namely
\begin{align}\label{sr2}
r_s=\frac{3\sqrt{3}M(1-\gamma)}{1-\kappa\eta^2}    
\end{align}
It is straightforward to notice that the LV parameter modifies the sizes of the black hole shadows. The Fig. (\ref{fig1}) depicts the shadow size, showing that the size decreases with the LV parameter and increases with the influence of the global monopole term. Additionally, we depict the shadow radius versus the parameters $\gamma$ and $\eta$ in Fig. (\ref{fig2}).

We can use observational data from Event Horizon Telescope (EHT) related to black hole shadow of Sagittarius A* (Sgr A*), a supermassive black hole at the center of the Milky Way galaxy, to estimate bounds for the parameters $\gamma$ and $\eta$. Firstly, let us define the radius of the observed shadow of Sgr A* as follows
\begin{align}
    R_{Sgr A*}=\frac{D\, \Theta_s}{2M_{Sgr A*}},
\end{align}
where $\Theta_s$ represents the angular size of the black hole shadow of  Sgr A* galactic center reported by EHT, $M_{Sgr A*}$ is the mass of Sgr A* and $D$ is the distance to Sgr A*. According to EHT data based on estimative from the Keck and Very Large Telescope Interferometer (VLTI) instruments, one has $\Theta_s=(48.7\pm 7)\,\mu a s$, $M_{Sgr A*}=4\times 10^{6}M_{\odot}$ and $D=8\,kpc$ \cite{EventHorizonTelescope:2022wkp, EventHorizonTelescope:2022exc, EventHorizonTelescope:2022xqj}. With this data in hands, we obtain the shadow radius of Sgr A* in mass units given by
\begin{align}
  R_{Sgr A^{*}}= (5.0\,\pm 0.5)M_{Sgr A*}.
\end{align}
In order to make a comparison between the observational and theoretical values, let us consider some values for the parameters $\gamma$ and $\eta$. Firstly, from the shadow radius (\ref{sr2}) with $\gamma=\eta=0$, we get the theoretical result
\begin{align}
    r_s=5.196M_{Sgr A*}.
\end{align}
If we set $\gamma=0.10$ and $\kappa\eta^2=0.15$, we find the following result
\begin{align}
    r_s=5.50M_{Sgr A*},
\end{align}
which is the upper limit of the shadow radius of Sgr A*. On the other hand, by setting $\gamma=0.25$ and $\kappa\eta^2=0.15$, one finds
\begin{align}
    r_s=4.58M_{Sgr A*},
\end{align}
which is the lower limit of the shadow radius of Sgr A*. In general, to agree with the observational data of the shadow radius of Sgr A*, the parameters $\gamma$ and $\eta$ must satisfy the following relation
\begin{align}
    0.867 \lesssim \frac{1-\gamma}{1-\kappa\eta^2}   \lesssim 1.059,
\end{align}
which imposes bounds on the parameters $\gamma$ and $\kappa\eta^2$.


\section{Scattering and Absorption}\label{s5}

\subsection{Probe real scalar field}
Once the global monopole black hole geometry is defined, we can study the behavior of a scalar field in this geometry. The action for a  massless scalar field  minimally coupled to gravity looks reads
\begin{align}
    S_{CE}=-\frac{1}{2}\int d^4x\sqrt{-g}\nabla_\mu\psi\nabla^\mu\psi.
\end{align}
Here we assume that $\psi$ is a probe field, so that we can neglect any back-reaction. By varying this action with respect to the field $\psi$, we obtain its equation of motion
\begin{align}\label{ce}
    \nabla_\mu\nabla^\mu\psi=\partial_\mu (\sqrt{-g} g^{\mu\nu}\partial_\nu \psi)=0.
\end{align}
For the metric (\ref{metric}), one has $\sqrt{-g}=r^2\sin\theta$, so, Eq. (\ref{ce}) reads
\begin{align}\label{eqce}
    -\frac{r^2}{F}\frac{\partial^2\psi}{\partial t^2}+\frac{\partial}{\partial r}\bigg(r^2 F \frac{\partial\psi}{\partial r}\bigg)+\frac{1}{\sin\theta}\frac{\partial}{\partial \theta}\bigg(\sin\theta\frac{\partial\psi}{\partial \theta}\bigg)+\frac{1}{\sin\theta}\frac{\partial^2\psi}{\partial \phi^2}=0.
\end{align}

Let us perform a separation of variables through adopting the following ansatz:
\begin{align}
    \psi(t,r,\theta,\phi)=\frac{\mathcal{R}(r)}{r}Y_{lm}(\theta,\phi)e^{-i\omega t}.
\end{align}
Afterwards, the Eq. (\ref{eqce}) becomes
\begin{align}\label{ece}
    F\frac{d}{dr}\bigg(F\frac{d\mathcal{R}}{dr}\bigg)+[\omega^2-\mathcal{V}_{eff}(r)]\mathcal{R}=0,
\end{align}
where $V_{eff}$ is the effective potential
\begin{align}
  \mathcal{V}_{eff}(r)=\frac{F}{r}\frac{dF}{dr} + \frac{F\  m(m+1)}{r^2}. 
\end{align}
Using $F(r)$ defined by (\ref{sol2}) for the case of the zero cosmological constant, the effective potential can be written as
\begin{align}
   \mathcal{V}_{eff}(r)=\bigg(\frac{1-\kappa\eta^2}{1-\gamma}-\frac{2M}{r}\bigg)\bigg(\frac{2M}{r^3}+\frac{m(m+1)}{r^2}\bigg).
\end{align}
This effective potential can be used in further calculations.

\subsection{Scattering}
In this section, we will calculate the differential scattering cross-section for the global monopole black hole using the partial wave method in the low-frequency regime (see f.e. \cite{Anacleto:2020zhp, Anacleto:2019tdj,Pitelli:2017bgx, Anacleto:2017kmg, Anacleto:2022shk}). For our purpose, we express Eq. (\ref{ece}) in the
Schr\"{o}dinger form using the transformation $\chi(r)=\sqrt{F(r)}\mathcal{R}(r)$. Thus, we have
\begin{align}\label{echi}
    \frac{d^2\chi(r)}{dr^2}+U(r)\chi(r)=0,
\end{align}
where
\begin{align}
    U(r)=\frac{F^{\prime 2}}{4F^2}-\frac{F^{\prime \prime}}{2F}+\frac{(\omega^2-\mathcal{V}_{eff})}{F^2}.
\end{align}

One can expand (\ref{echi}) in power series in $1/r$, so that we have
\begin{align}
U(r)&=\frac{(1-\gamma)^2 \omega ^2}{\left(1-\kappa\eta ^2\right)^2}-\frac{4 (1-\gamma)^3 M \omega ^2}{r \left(1-\kappa\eta ^2\right)^3}+\frac{12\ell^2}{r^2}\nonumber\\&+\frac{2 (1-\gamma)^2 M \bigg[3 m (m+1) \left(1-\kappa\eta ^2\right)^3-16 (1-\gamma)^3 M^2 \omega ^2\bigg]}{r^3 \left(1-\kappa\eta ^2\right)^5}+\cdots,
\end{align}
with
\begin{align}
    \ell^2=\frac{M^2\omega^2(1-\gamma)^4}{\left(1-\kappa\eta ^2\right)^4}-\frac{m(m+1)(1-\gamma)}{12\left(1-\kappa\eta ^2 \right)}.
\end{align}

The above $\ell^2$ is defined as the change in the coefficient of $1/r^2$ (including only contributions involving the quantities $\ell$ and $\omega$) arising after performing the power series expansion in $1/r$ in (\ref{echi}). Note that as $r \rightarrow \infty$, the potential tends to zero, $U(r)\rightarrow 0$, reproducing the asymptotic value. Thus, knowing the phase shift allows one to obtain the scattering amplitude, which has the following partial wave representation.
\begin{align}
    f(\theta)=\frac{1}{2i\omega}\sum_{m=0}^\infty (2m+1)(e^{2i\delta_m}-1)P_m(\cos\theta),
\end{align}
and the differential scattering cross section can be computed by the expression
\begin{align}
  \frac{d\sigma}{d\Omega}=\vert f(\theta)\vert^2.  
\end{align}
To obtain the phase shift, we use
\begin{align}
    \delta_l=2(m-\ell).
\end{align}
Thus, in the limit $m \rightarrow 0$, we obtain
\begin{align}
 \delta_0=-\frac{2M\omega\,(1-\gamma)^2}{(1-\kappa\eta^2)^2}.   
\end{align}
Note that in the limit $l \rightarrow 0$, the phase shifts tend towards nonzero terms, naturally leading to a correct result for the differential cross-section in the small angle limit. At this point, we should note that the representation is poorly convergent, which makes the task of performing the sum of the series rather difficult. Consequently, we are led to the problem where an infinite number of Legendre polynomials are required to obtain divergences in $\theta=0$. In order to circumvent such a problem, we utilize a reduced series, which is less
divergent in $\theta=0$, explicitly,
\begin{align}
(1-\cos\theta)^m\,f(\theta)=\sum_{m} a_m\, P_m(\cos\theta).  
\end{align}
Then, to determine the differential scattering cross section, we can use the following equation 
\begin{align}
\frac{d\sigma}{d\Omega}=\bigg\vert\frac{1}{2i \omega}  \sum_{m} (2m+1)(e^{2i\delta_m}-1)\frac{P_m(\cos\theta)}{1-cos\theta}\bigg\vert,  
\end{align}
so that the cross section for $m$ ($m=0$) is given by
\begin{align}
    \frac{d\sigma}{d\Omega}=\frac{4\delta_0^2}{\omega^2\theta^4}=\frac{16 M^2 (1-\gamma)^4}{(1-\kappa\eta^2)^4\theta^4}
\end{align}
As we can verify, the differential cross section is decreased by the LV effect, while it is increased by the monopole effects. Additionally, one can employ another path to obtain the same phase shift, which is through the Born approximation formula, namely
\begin{align}
    \delta_l=\frac{\omega}{2}\int_0^\infty dr r^2 J_l^2(\omega r) u(r).
\end{align}
where $J_l(\omega r)$ stand for the spherical Bessel functions of the first kind and $u(r)$ is the potential given by
\begin{align}
u(r)=\frac{2 (1-\gamma)^2 M \bigg[3 m (m+1) \left(1-\kappa\eta ^2\right)^3-16 (1-\gamma)^3 M^2 \omega ^2\bigg]}{r^3 \left(1-\kappa\eta ^2\right)^5}+\cdots.    
\end{align}
Now, let turn our attention to the phenomena of absorption ahead.

\subsection{Absorption}
Now, let us determine the absorption cross section for the black hole in the low-frequency limit. A result well known in quantum mechanics is the total absorption cross section, which can be computed by means of the following relation
\begin{align}
   \sigma_{abs}&=\frac{\pi}{\omega^2}\sum_{m=0}^{\infty}(2m+1)(1-e^{2i\delta_m})=\frac{4\pi}{\omega^2}\sum_{m=0}^{\infty}(2m+1)\sin^2{\delta_m}\nonumber\\&=\frac{4\pi}{\omega^2}\bigg[\sin^2{\delta_0}+\sum_{m=1}^{\infty}(2m+1)\sin^2{\delta_m}\bigg]. 
\end{align}
Thereby, we obtain in the limit $\omega\rightarrow 0$ the following total absorption cross section for the phase shift 
\begin{align}
    \sigma_{abs}=\frac{4\pi\delta_0^2}{\omega^2}=\frac{16\pi M^2 (1-\gamma)^4}{(1-\kappa\eta^2)^4}.
\end{align}
We can still write in terms of the area of the event horizon of the Schwarzschild black hole $\mathcal{A}_{Sch}=16\pi M^2$, thereby arriving at
\begin{align}
\sigma_{abs}=\frac{\mathcal{A}_{Sch}(1-\gamma)^2}{(1-\kappa\eta^2)^2}.    
\end{align}
One notes that there is a contribution of both the LV parameter and of the
monopole term. Absorption decreases as we increase $\gamma$, while it increases as $\eta$ grows. On the other hand, when $\gamma=0$, we recover the result obtained for a black hole with global monopole \cite{Anacleto:2017kmg}, and when $\gamma=\eta=0$, we meet the result obtained in \cite{Sanchez:1976xm} for the Schwarzschild black hole. It is noteworthy that the result agrees with the universality property of the absorption cross section which is always proportional to the area of the event horizon at the low frequency limit.

\section{Greybody factor}
Around the black hole, spacetime is strongly curved, generating a gravitational barrier. In this context, the greybody factor (GF) plays an essential role by indicating the probability of a quantum field (Hawking radiation) escaping from this barrier. The maximum value of the greybody bound corresponds to the blackbody, being equal to one. In this section, we will focus on calculating the GF of massless scalar field for the black hole solution studied so far. To achieve such a purpose, we will employ the method adopted in \cite{Okyay:2021nnh, Kumaran:2023brp, Singh:2024nvx, Baruah:2025ifh, Heidari:2024bvd}. As a starting point for our analysis, let us rewrite Eq. (\ref{ece}) as follows
\begin{align}\label{ece1}
    \frac{d^2\mathcal{R}}{dr^{\ast 2}}+[\omega^2-\mathcal{V}_{eff}(r)]\mathcal{R}=0,
\end{align}
where we have defined the tortoise coordinate $\frac{dr^{\ast}}{dr}=\frac{1}{F}$ and $V_{eff}$ is the effective potential
\begin{align}\label{eff}
  \mathcal{V}_{eff}(r)=\frac{F}{r}\frac{dF}{dr} + \frac{F\  m(m+1)}{r^2}. 
\end{align}

With the effective potential at hand, we can use it to compute the GF of the scalar field. We achieve such a goal by employing the general semi-analytic bounds, enabling us to conduct a qualitative analysis of the results. Therefore, we define the transmission probability as follows
\begin{align}
 \sigma(\omega)=\mathrm{sech}^2\bigg(\int_{-\infty}^{\infty}\rho(r_{\ast}) dr_{\ast}\bigg)   
\end{align}
where
\begin{align}
\rho(r_{\ast})=\frac{1}{2h}\sqrt{\bigg(\frac{d h(r_{\ast})}{dr_{\ast}}\bigg)^2+(\omega^2-\mathcal{V}_{eff}(r)-h^2(r_{\ast}))^2}    
\end{align}
Above, $h(r_{\ast})$ represents a positive function satisfying $h(\infty)=h(-\infty)=w$
\begin{align}
\sigma(\omega)=\mathrm{sech}^2\bigg(\frac{1}{2\omega}\int_{r_H}^{\infty}V_{eff}(r_{\ast})\, dr_{\ast} \bigg)   
\end{align}

\begin{figure}[h]
\begin{center}
\hspace*{-3mm}
\begin{tabular}{ccc}
\includegraphics[scale=0.3]{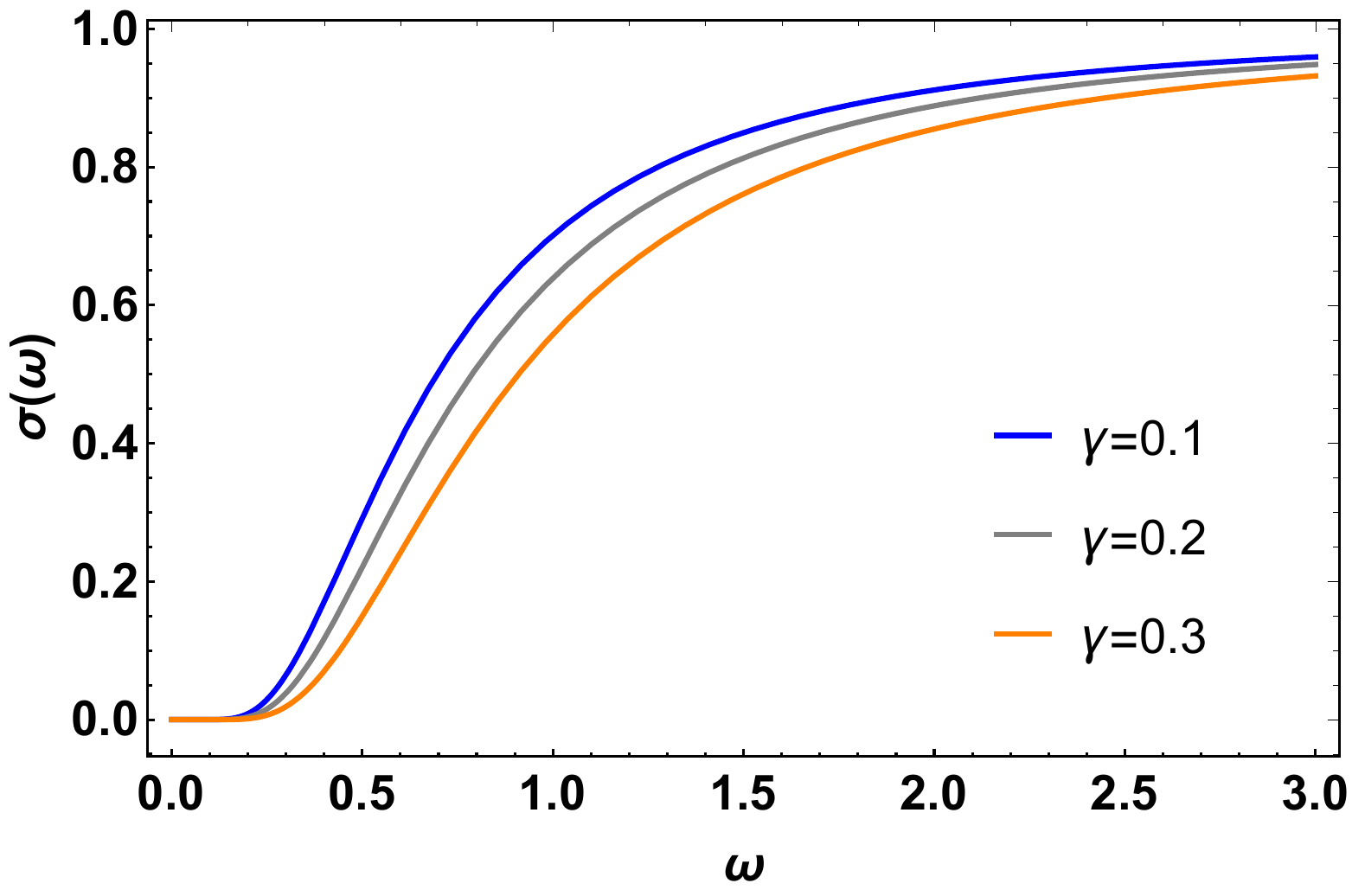} 
\includegraphics[scale=0.3]{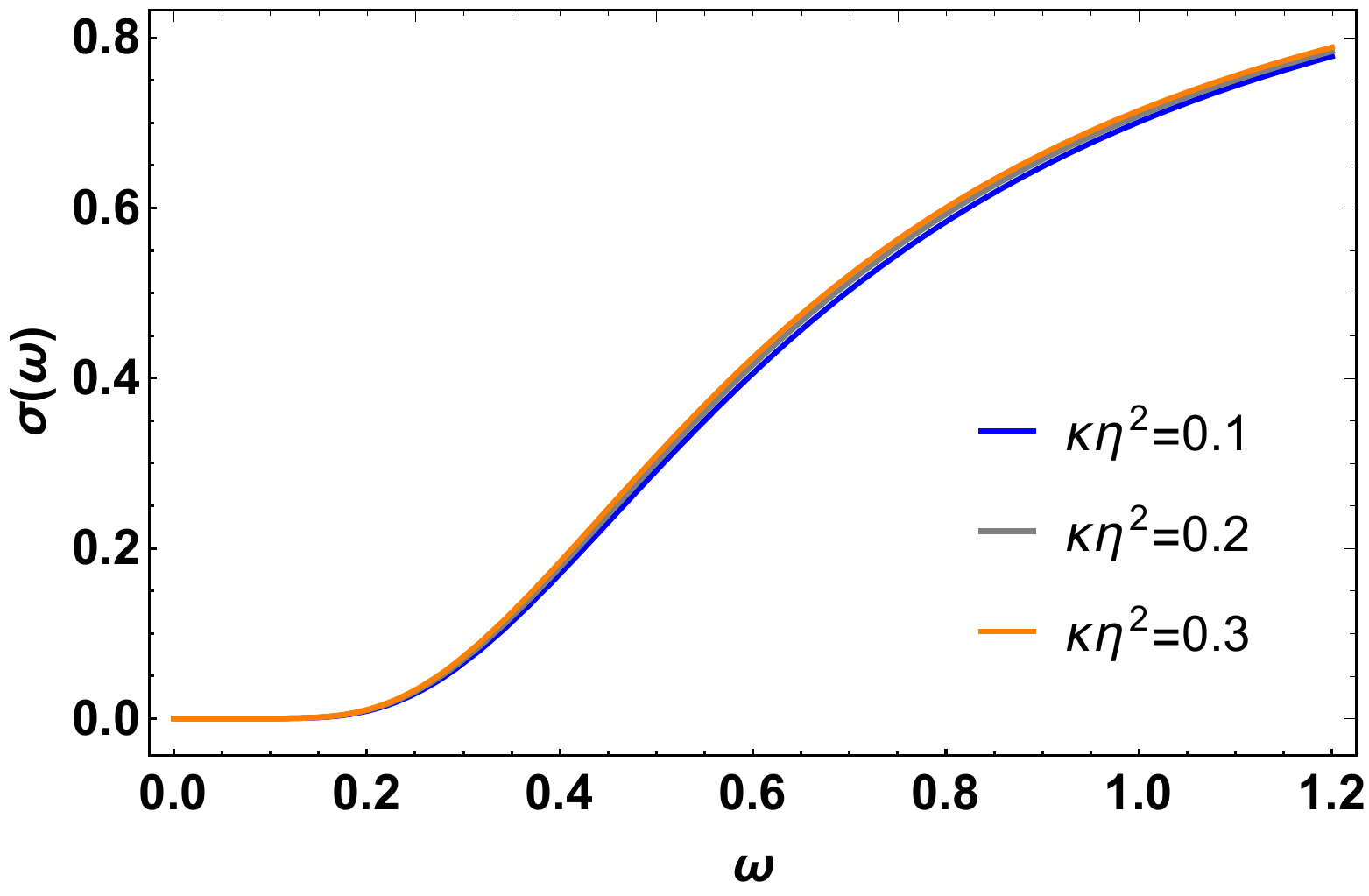}\\
(a)\hspace{8.5cm}(b)\\
\end{tabular}
\end{center}
\vspace{-0.5cm}
\caption{  The representation of the greybody factors as a function of $\omega$ by varying the parameters $\gamma$ for $\kappa\eta^2=0.1$ (a) and $\eta$ for $\gamma=0.1$ (b).
\label{fig3}}
\end{figure}

It is easy to see that the radial function $F(r)$ plays a pivotal role in determining the relationship between GFs and the effective potential. Then, the GF with the effective  potential (\ref{eff}) is given by 
\begin{align}
  \sigma(\omega)=\mathrm{sech}^2\bigg[\frac{1}{2\omega}\int_{r_H}^{\infty}\bigg(\frac{F^{\prime}}{r}+\frac{m(m+1)}{r^2}\bigg) dr \bigg]  
\end{align}
Using the function \ref{sol1}, we arrive at
\begin{align}
    \sigma=\mathrm{sech}^2\bigg[-\frac{1}{2\omega}\bigg(\frac{M}{r_h^2}+\frac{m(m+1)}{r_h}\bigg)\bigg]
\end{align}
The GF depends on both the LV parameter and on the global monopole term. We depict the behavior of GF in Fig.(\ref{fig3}). As we can see, when the frequency is at a lower level, we observe that radiation is reflected totally and there is no transmission. The tunneling effect occurs as the frequency increases so that the radiation can pass through the potential barrier. Besides, the greybody factor decreases with larger values of the LV parameter, while it slightly increase as we modify the global monopole term. In this case, we physically interpret this as follows: when the parameter $\gamma$ increases, the height of the potential barrier also increases, leading to a lower probability of transmission to the incident wave. 

\section{Final remarks}\label{s6}

In this work, one investigated a static spherically symmetric spacetime in the context of a gravity non-minimally coupled to the Kalb-Ramond field with non-zero VEV. Such a non-zero VEV induces the breaking of the Lorentz symmetry. As matter content, global monopole was considered. In this conjecture, for a suitable configuration of the KR field, we were able to obtain an analytical solution in the absence and presence of cosmological constant. We also calculated the Hawking temperature, showing the influence of LSV and the global monopole.

Some physical implications of these solutions were investigated by means of the results of classical experiments regarding the Solar System. Within this study, we concentrated on the advance of Mercury's perihelion and light deflection, establishing constraints for the symmetry-breaking energy scale of the global monopole, $\eta$. For the former, we estimated the parameter $\eta$ by the requirement that any additional perihelion advance from a global monopole does not exceed the observational uncertainty ($\pm 0.003^{\prime\prime}/\mbox{century}$) leads to the upper bound constraint $\eta< 8\times 10^{12}$ GeV. This means that if a global monopole would present in the Solar System, the symmetry‐breaking scale $\eta$ would have to be lower than roughly $8\times 10^{12}$ GeV in order to not produce a detectable extra precession of Mercury’s perihelion.

 Employing the method of partial waves, we examined the scattering and absorption of a massless scalar field using the solution obtained for the case of a zero cosmological constant. We showed that the cross-sections explicitly depend on $\gamma$ and $\eta$, as expected. In particular, we demonstrated that the absorption cross-section holds the universality property of being linearly dependent to the area of the Schwarzschild event horizon. 
 
 We also obtained an exact expression for the greybody factor, which depends explicitly on both the LSB parameter and the global monopole charge. The effect of LSB is to increase the greybody factor as $\gamma$ lowers, while the global monopole tends to decrease the greybody factor as $\eta$ lowers.

A forthcoming study could involve investigating spinning black holes and analyzing the impact of the LSB parameter and the monopole charge on these solutions. We plan to conduct this study in the future.

\section*{Acknowledgments}
\hspace{0.5cm} The authors thank the Funda\c{c}\~{a}o Cearense de Apoio ao Desenvolvimento Cient\'{i}fico e Tecnol\'{o}gico (FUNCAP), the Coordena\c{c}\~{a}o de Aperfei\c{c}oamento de Pessoal de N\'{i}vel Superior (CAPES), and the Conselho Nacional de Desenvolvimento Cient\'{i}fico e Tecnol\'{o}gico (CNPq).  Fernando M. Belchior thanks the Departmento de F\'isica da Universidade Federal da Para\'iba - UFPB for the kind hospitality and has been partially supported by CNPq grant No. 161092/2021-7. Roberto V. Maluf thanks the CNPq for grant no. 200879/2022-7. Albert Yu. Petrov and  Paulo J. Porf\'irio would like to acknowledge the CNPq, respectively for grant No. 303777/2023-0 and grant No. 307628/2022-1.

\end{document}